\documentclass[prl,showpacs,twocolumn,preprintnumbers]{revtex4-1}

\usepackage{mathptmx,amssymb}
\usepackage{amsmath}
\usepackage{graphicx}

\begin{document}

\title{Sudden chain energy transfer events in vibrated granular media}

\author{Nicol\'as Rivas$^1$,  
	Suomi Ponce$^1$,
	Basile Gallet$^3$,
        Dino Risso$^2$, 
        Rodrigo Soto$^1$,
        Patricio Cordero$^1$,
	Nicol\'as Mujica$^1$}
\affiliation{$^1$Departamento de F\'{\i}sica, FCFM, Universidad de Chile,
        Santiago, Chile\\
        $^2$ Departamento de F\'{\i}sica, Universidad
        del  B\'{\i}o-B\'{\i}o, Concepci\'on,  Chile \\
	$^3$ Laboratoire de Physique Statistique, CNRS, Ecole Normale Sup\'erieure,
	Paris, France}


\begin{abstract} 
In a mixture of two species of grains of equal size but
different mass, placed in a vertically vibrated shallow box, there is
spontaneous segregation.  Once the system is at least partly segregated and 
clusters of the heavy particles have formed, there are sudden peaks of the 
horizontal kinetic energy of the heavy particles, that is otherwise small. 
Together with the energy peaks the clusters rapidly expand and the segregation
is partially lost. The process repeats once segregation has taken place again.
Depending on the experimental or numerical parameters,
the energy bursts can occur either randomly or with some regularity in time.
An explanation for these events is provided based on the existence of a fixed
point for an isolated particle bouncing with only vertical motion. 
The horizontal energy peaks occur when the energy stored
in the vertical motion is partly transferred into horizontal energy through a
chain reaction of collisions between heavy particles.   
A necessary condition for the observed regularity of the events is 
that chain reactions involve most of the heavy particles.
 \end{abstract}

\maketitle

{\em Introduction.} Granular media, when externally excited, can have a
fluid-like behavior showing many flow regimes resembling those of molecular
fluids even though there are strong differences due to the energy
dissipation at grain collisions \cite{cinco}.  Fluidized granular media
remain in a far from equilibrium regime, presenting phenomena that makes
difficult the development of hydrodynamic models with quantitative
predictive power.  Among other specific phenomena, we remark absence of
scale separation \cite{scale-sep}, lack of energy equipartition
\cite{non-equipartition}, violation of the fluctuation-dissipation relations
\cite{fluctdis}, non thermodynamic fluxes \cite{NonFourier}, spontaneous
development of inhomogeneities \cite{inhomogeneous}, segregation
\cite{Kudrolli}, and localized surface structures \cite{oscillons}.  
A careful description of these and other
phenomena are crucial to develop models that describe the collective
behavior of granular fluids.
 
Fluidized granular media in a shallow geometry (quasi two dimensional) has attracted
attention because it allows for a detailed analysis of both the collective
behavior and the motion of individual grains
\cite{98olafsen,ccdhmrv08,shallowhoriz1,shallowhoriz2}. 
The possibility of quantifying the system's dynamics
at both scales may help building a mathematical model for the collective
dynamics of granular media. Placing
monodisperse inelastic spheres in a vertically vibrated shallow box of
height less than two particles' diameters, a particular phase separation
takes place: grains form solid-like regions surrounded by fluid-like ones,
having high contrasts in density, local order and granular temperature \cite{98olafsen}. 
This phase separation 
is driven by the negative compressibility of the effective two dimensional
fluid \cite{ccdhmrv08}. For shallow systems the horizontal kinetic energy
of the grains can be quite different from the vertical kinetic energy.

Granular matter is usually polydisperse, with grains differing in mass, shape, size or mechanical properties. It is known that a mixture of
two types of grains differing in some of these properties
can mix or segregate when externally excited~\cite{Kudrolli}. 

In this letter we report an experimental and numerical study of a phenomenon
that takes place when two particle species of equal size but different mass
are put in a vertically vibrated shallow box. The focus of
the present letter is not on segregation, but on a quite peculiar phenomenon 
that takes place once the species have segregated:
the horizontal energy of the system has sudden and brief peaks, together with 
fast expansions of the regions rich in heavy particles. This is what we call {\it chain energy transfer events}. The characterization of these events shows a direct link between the individual particle dynamics and the observed collective behavior. 

In experiments, the two species do not only differ in their mass but,
being made of different materials, they also differ in their mechanical
properties. Numerical simulations allow then to identify the key elements that
generate this phenomenon.  Simulating the simple hard sphere model it becomes
evident that the mass difference is the key parameter that controls the occurrence of energy peaks and
sudden expansions; the differences in mechanical properties only change them
quantitatively. 
Simulations give information on the particle's vertical motion, helping to complete the description of the phenomenon.

{\em System configuration.} Two species of spherical grains of the same
diameter $\sigma$ but different mass are placed in a
square shallow box as described below.  The lighter/heavier particles will
be called $L\,\,/\,\,H$.

In the experiments, we use a mixture of $N_H = 170$ stainless steel (density
$7.8\,\text{g/cm$^3$}$) and $N_L = 291$ polyamide (density
$1.14\,\text{g/cm$^3$}$) particles, both of $\sigma=3$ mm.  The shallow box
has dimensions $L_x/\sigma = L_y/\sigma = 33.33 \pm 0.03$ and $L_z/\sigma =
1.813\pm0.004$.  The box consists of two $10$ mm thick ITO coated glass plates separated
by a square steel frame.  It is fixed to a base, which supports an array of
high intensity LEDS.  This whole setup is forced sinusoidally with an
electromechanical vibrator.  An accelerometer is fixed to the base of the cell, which allows the measurement of the imposed forcing amplitude. Top view images are obtained with a high speed
camera. Particle positions and velocities can be determined at
sub-pixel accuracy.  The vibration
frequency is fixed to $f =\omega/2\pi = 1/T=100$ Hz. Two values of the
vibration amplitude are used, $A/\sigma=0.036$ and $A/\sigma=0.045$.  

In the simulations, we use instantaneous collisions that are characterized
by restitution and friction coefficients, which are velocity independent 
and different for both kind of particles, but the same for particle-particle and particle-wall collisions. 
An event driven algorithm is used \cite{eventdriven}. 
The shallow
box oscillates vertically
with a biparabolic waveform.  All simulations start with a random initial mixing. 

Here, we report two types of simulations: 
The first type (S1) corresponds to simulations performed at higher densities than the experiment. Its purpose is to show that the chain energy transfer events are robust. In these simulations, the box has flat bottom and top walls and horizontal periodic boundary conditions. We use the following parameters: mass ratio between a heavy and a light particle $m_H/m_L = 10$, box size
$L_x/\sigma = L_y/\sigma = 40$, $L_z/\sigma = 1.82$, normal and tangential
restitution coefficients $\alpha_L = \alpha_H = 0.8$, static and dynamic friction
coefficients $\mu_s = 0.3$ and $\mu_d = 0.15$, amplitude and frequency of
vibration $A/\sigma = 0.15$, $\omega \sqrt{\sigma/g} = 7$.  The number of
particles are $N_L = 1000$ and $N_H = 500$, giving the filling fraction $\rho = N\pi\sigma^2/4L_xL_y = 0.736$. Other box sizes, filling fractions and heavy particle  concentration were simulated with results similar to the ones we are presenting.
 
The second type of simulations (S2) are performed to find which experimental imperfections are relevant for the particular observations (for example, the specific time and energy scales). Possible imperfections are the plate surface roughness and small curvature. The latter turns out to be more important than the former. In order to consider this, we use a slightly parabolic bottom plate, with a variation of height between the bottom and flat top plates of only $1\%$.  In these simulations we use lateral hard walls. 
Comparing S2 with S1 we learn
which are the essential ingredients of the phenomenon and which aspects only change quantitatively the observations. For S2 simulations, we use the following parameters:  $m_H/m_L = 7.8$,  $L_x/\sigma = L_y/\sigma = 33$, $L_z/\sigma = 1.78$, $\alpha_L = 0.97$, $\alpha_H = 0.94$, $\mu_s = 0.1$ and $\mu_d = 0.05$, $\omega \sqrt{\sigma/g} = 11.0$ and $A/\sigma = 0.055$ or $A/\sigma = 0.062$.  The number of $L$ and $H$ particles are the same as in experiments, with $\rho = N\pi\sigma^2/4L_xL_y = 0.44$.

Unless explicitly stated, the following descriptions refer to both simulations and
experiments. Shortly after starting from a homogeneous configuration segregation 
between the two species takes place and many small dense clusters of $H$'s appear.  
Later on, the clusters coalesce and tend to have some $L$'s in their bulk.
From a top view the $H$'s
appear as if they were standing still, while the $L$'s outside the clusters
show a significant horizontal agitation.  The external pressure exerted by
the $L$'s leads the $H$'s to form dense clusters~\cite{videos}.
The coalescence process, that is normally slow if 
the box is perfectly flat (S1), is sped up if there is a small curvature of the 
plates, coming from slight imperfections of the experimental setup, or imposed explicitly in a numerical simulation (S2), with the resulting cluster located at the center.
In either case, the subsequent description of the phenomenon is independent
of any small curvature. It only depends on the density of the cluster of the $H$'s and on the number and distribution of light particles which remained trapped inside it.

\begin{figure}[t!] 
\begin{center}
\includegraphics[width=.9\columnwidth]{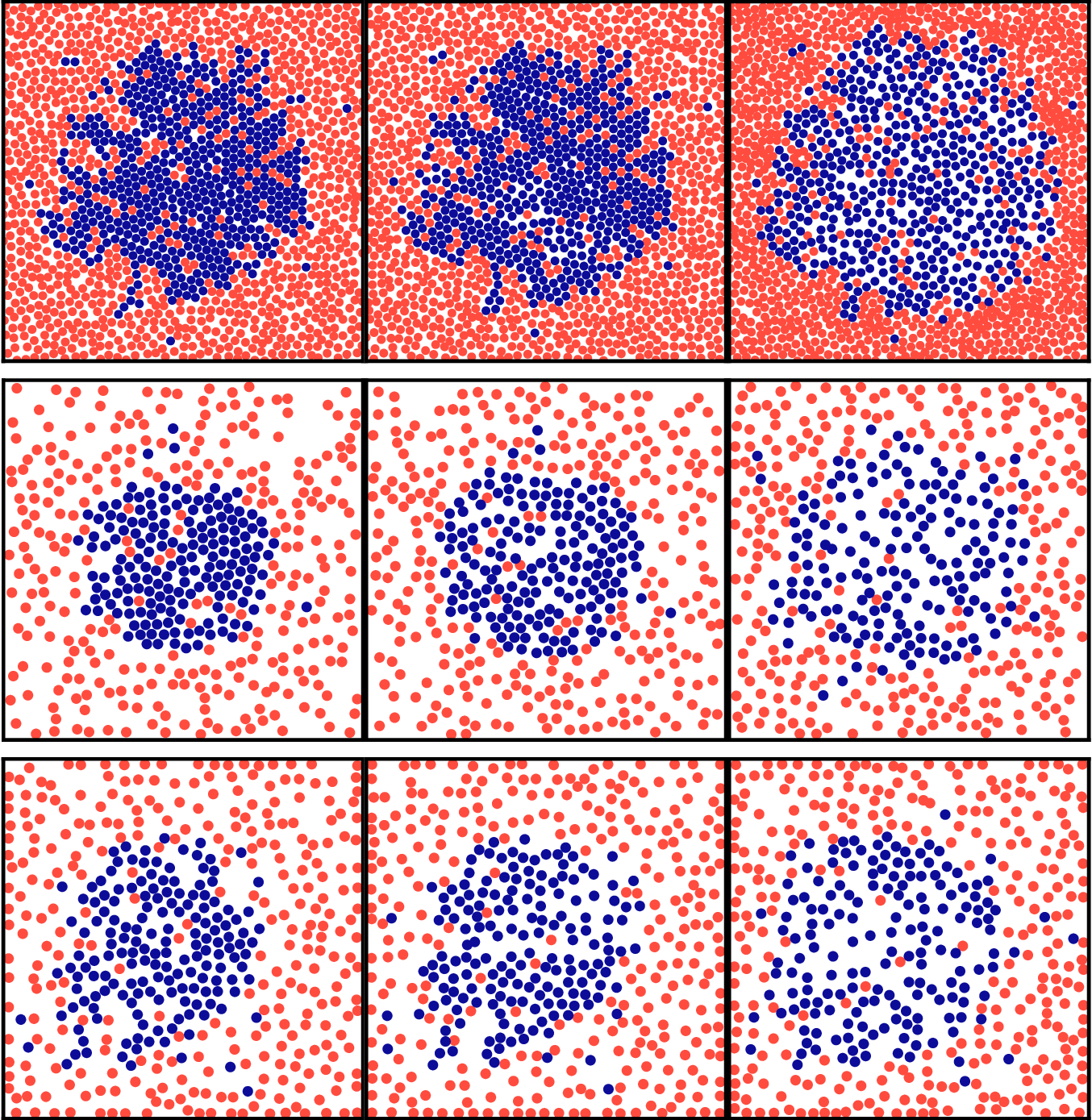}
\end{center} 

\caption{(Color online) Sequence of top view configurations for simulations S1 (top), S2 (middle) 
and experiments (bottom), showing the development of an energy peak. 
Heavy ($H$) particles are presented in blue and light ($L$) ones in
red.  First, the  system reaches partial segregation.  Then
the beginning of the event is seen as a small region of lower density within
the cluster.  Next, the low density region has propagated to practically all
the cluster.  Later (not shown) the system comes back to a state similar to the
one before the expansion.}
\label{fig:ExpFotos}
\end{figure}

 \begin{figure}[t!]
\begin{center}
 \includegraphics[width=.92\columnwidth]{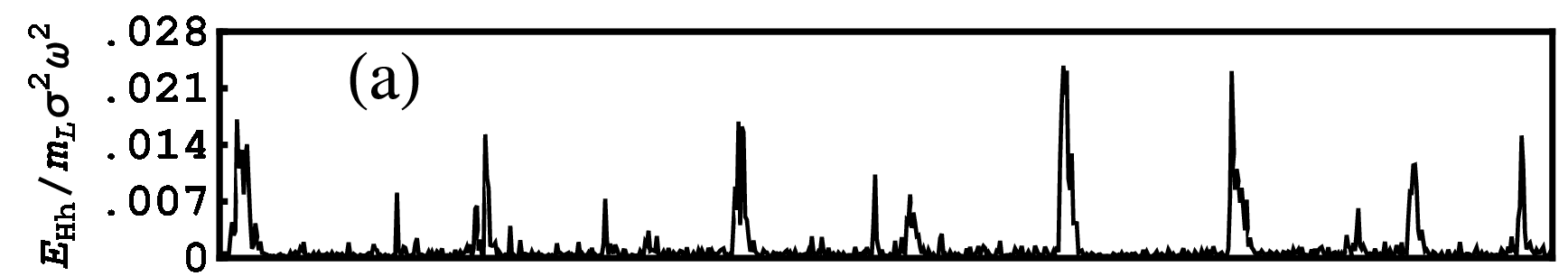}
 \includegraphics[width=.92\columnwidth]{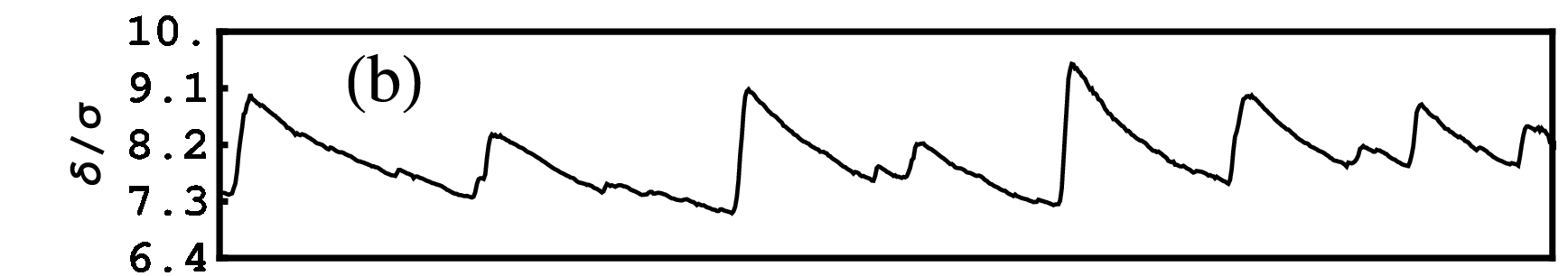}
 \includegraphics[width=.92\columnwidth]{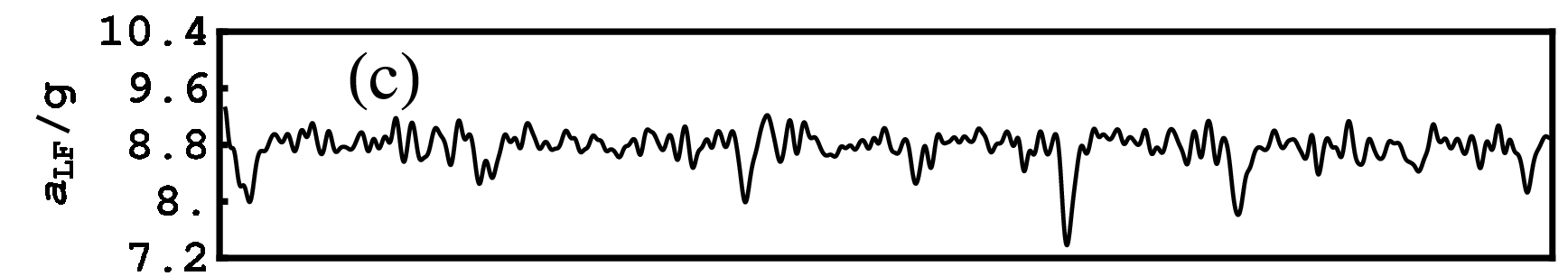}
 \includegraphics[width=.92\columnwidth]{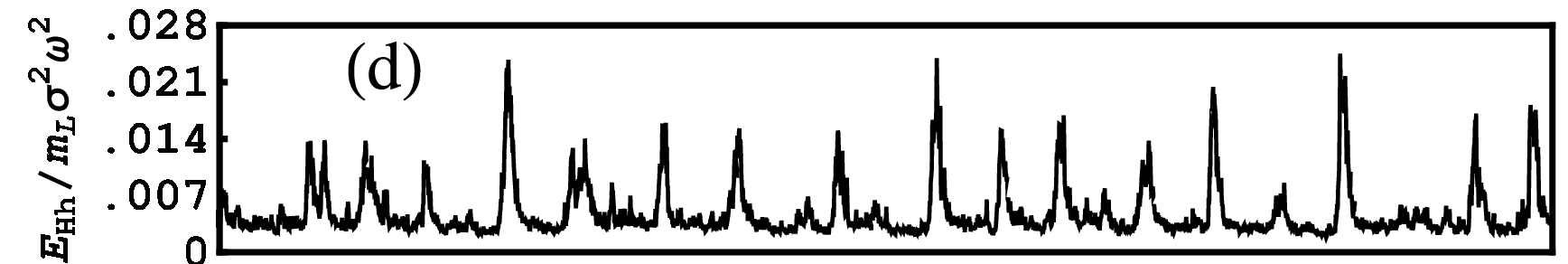}
 \includegraphics[width=.92\columnwidth]{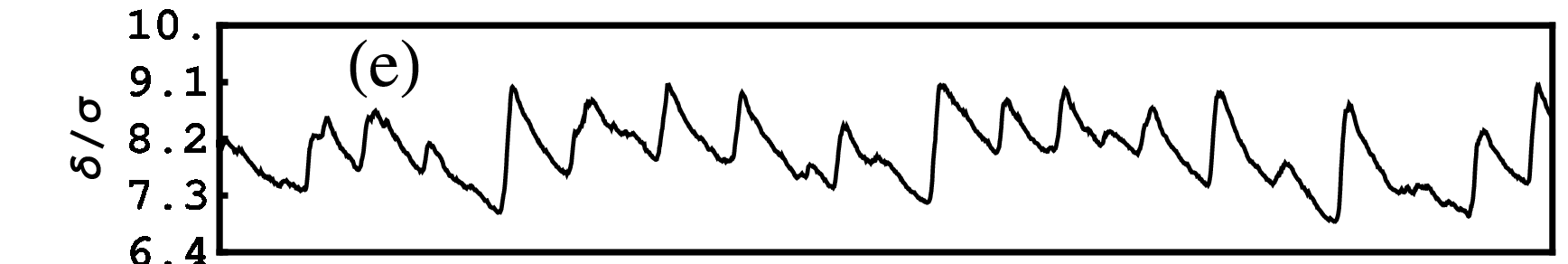}
 \includegraphics[width=.92\columnwidth]{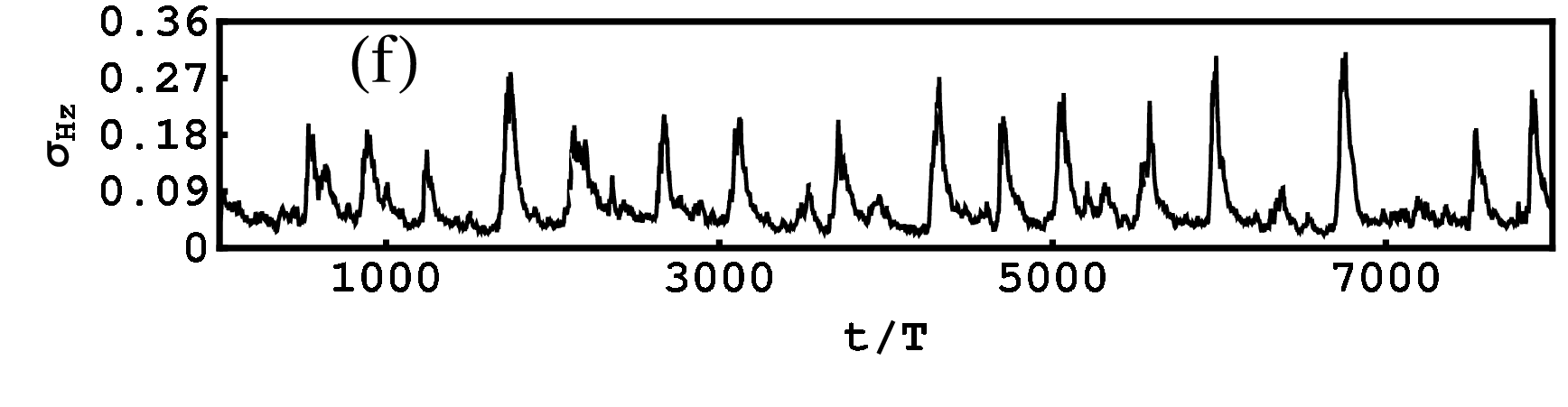}
\end{center}

\caption{Time evolution of $E_{Hh}$ (a),  
$\delta$ (b) and  $a_{\rm LF}$ (c) 
for an experiment with  $A/\sigma = 0.045$. $E_{Hh}$ (d), $\delta$ (e) and $\sigma_{\rm Hz}$ (f) for a simulation S1 and $A/\sigma = 0.062$ .}
\label{fig:EnergiaHorizontalB}
\end{figure}

{\em The chain events.} Once there is at least one cluster there is a sudden
phenomenon of short duration.  
Figure~\ref{fig:ExpFotos} shows typical temporal sequences of the grain positions when an event takes place,
for simulations S1, S2, and experiments.  A chain reaction begins as an abrupt increase
in the horizontal agitation of the $H$'s in a small region of one of the clusters,
implying a local expansion followed by a fast propagation of the horizontal
agitation to a larger zone.  After a short while the horizontal
agitation of the $H$'s quickly decays, recovering its original value.  Next
the $L$-$H$ collisions slowly compress the cluster again, eventually
recovering the original density. Both cluster size and horizontal agitation,
computed as mean quadratic radius $\delta$ and the average (per particle) horizontal energy of the
$H$'s, $E_{Hh}$, respectively, are shown in  Fig. \ref{fig:EnergiaHorizontalB} for a sequence of several chain events in experiments and S2 simulations. Together with the fast expansions of 
the cluster there are energy peaks. Later, $\delta$ slowly decreases until the 
next energy burst takes place. Indeed, after the expansion, the dynamics of $\delta$ is much slower than that 
of the energies, because the former is related to a conserved quantity (the mass), 
while energy is not conserved and evolves in fast temporal scales.

Simulations show that the system is highly anisotropic, with vertical
energies much larger than the horizontal ones for both species. 
Before an energy peak the horizontal agitation of the $L$'s is higher
than for the $H$'s.  The reason is that the $L$'s are having collisions
with particles of both types and, in particular, the collisions with the
$H$'s keep them excited both vertically and horizontally.  On the other hand
most $H$'s are at their {\em fixed point}, with almost all their energy 
concentrated in the vertical
motion.  The abrupt change of $E_{Hh}$  is accompanied by
simultaneous but weaker changes in the other computed energies.  This
behavior is generic.

Under certain conditions, a single grain bouncing in a shallow box 
vibrating at a frequency $\omega$
evolves towards a fixed point: the bouncing
movement becomes periodic with the same frequency $\omega$ and, because of
the friction with the horizontal walls, the particle does not move
horizontally neither does it rotate.  Namely all its kinetic energy is
vertical energy.  The role of the fixed point in the behavior of the whole
system is evident by looking at side view videos of 
experiments and simulations~\cite{videos}.  It
is seen that the $H$'s are bouncing in phase as practically one solid layer:
they are collectively trapped at the fixed point very close to each other. 
In fact, the fixed point can be easily computed in absence of gravity (a
good approximation in this case) with a resulting vertical energy for the
heavy particles in the simulations of $0.32$ $m_L(\sigma\omega)^2$, that
fully agrees with the value of $E_{Hv}$ before and after the energy-peak
event, confirming that most $H$'s are close
to the fixed point.  Hence, the collisions among heavy particles
create almost no horizontal agitation.  It is this coherent movement that is
destroyed when an energy-peak event takes place.

To check that the picture described above is correct we have measured in
simulations the standard deviation $\sigma_{\rm Hz}$ of the stroboscopic height
of the $H$'s when the box is at its lowest position, shown in Fig. \ref{fig:EnergiaHorizontalB}(f).  
$\sigma_{\rm Hz}$ is almost zero before the energy burst and it jumps an
order of magnitude during the event.  It recovers its typical
small value together with $E_{Hh}$, indicating that the $H$'s are again
moving in phase. The cross correlation function between $E_{Hh}$ and
$\sigma_{Hz}$ shows a clear maximum centered at null time delay,
confirming that the energy bursts are accompanied by a massive desynchronization of the
$H$'s. 
Experimentally, this sudden desynchronization is observed through $a_{LF}$, which is the envelope of the acceleration signal including particle-plate collisions, obtained by low-pass frequency filtering. The main contribution to $a_{\rm LF}$ comes from the simultaneous
$H$ particle-plate collisions, which therefore exert a large force on the
plate.  The negative peaks, that occur at the same instants than the energy peaks (see Fig. \ref{fig:EnergiaHorizontalB}(c)), correspond to the absence of coherent
collisions due to the dephasing of the $H$'s and the decrease in their vertical kinetic energy.

In summary, at the start of the dynamics the horizontal energy $E_{Hh}$ of
the $H$'s begins to decrease, reaching a small asymptotic value because
most of the $H$'s are approaching the fixed point.  Particles $L$ produce
only small perturbations to the heavier ones but, due to the mass
difference, the $H$'s can easily take the $L$'s off their fixed point
getting significant horizontal energy.  As a result, the $L$'s very seldom
reach their fixed point and if they do so, it is for a short time.  The
permanent collisions with the $L$'s that some of the $H$'s suffer,
allow for a residual non vanishing horizontal energy $E_{Hh}$.  The clusters
of $H$'s slowly evolve to higher densities.  At one moment, an $H$
may collide with an $L$ in such a way that the $H$ gets a large horizontal
velocity. The heavy particle then hits a neighboring energetic $H$ triggering a chain
reaction of collisions among neighboring $H$'s rapidly propagating within
the cluster implying a sharp peak in $E_{Hh}$.  
The main energy source for
the increase of $E_{Hh}$ is the vertical energy $E_{Hv}$ which correspondingly decreases
during the energy burst.  The chain energy transfer induces that the $E_{Lh}$ and $E_{Lv}$
follow a similar pattern to a smaller extent.
This sudden chain energy transfer from the vertical to the horizontal motion 
is responsible for the observed fast increase of the cluster radius $\delta$.

\begin{figure}[t!] 
\begin{center}
\includegraphics[width=\columnwidth]{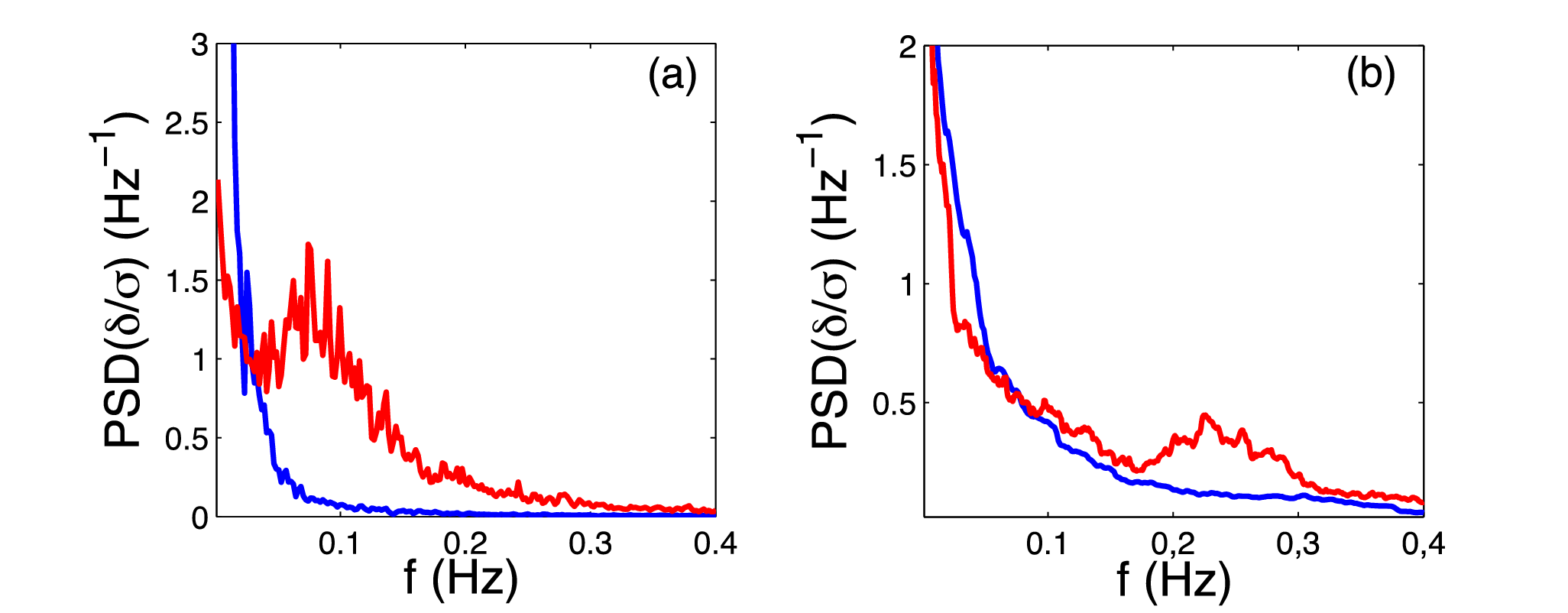}

\end{center} 

\caption{(Color online) Power spectrum density of $\delta/\sigma$. (a) Experimental results for $A/\sigma = 0.036$ and $A/\sigma =0.045$ (blue and red curves respectively); (b) Simulational results (S2) for $A/\sigma = 0.055$ and $A/\sigma =0.062$ (blue and red curves respectively).  }
\label{fig:PSD}
\end{figure}

{\em Statistical analysis.} 
The time lapses between successive $E_{Hh}$ peaks and their intensities show different
degrees of regularity; in some cases the peaks seem almost periodic. 
Fig. \ref{fig:PSD} presents the power spectral densities of $\delta/\sigma$ for both experiments and simulations S2. Two distinct regimes are observed: for low amplitude there is no periodicity in the signal, while for $A/\sigma=0.045$ for experiments ($A/\sigma=0.062$ for S2 simulations) a clear maximum, although broad, is observed. It is centered at a period $\approx1000T$ ($\approx400T$ for simulations), which is consistent with
the time series in Fig.  \ref{fig:EnergiaHorizontalB}b (Fig. \ref{fig:EnergiaHorizontalB}e). Depending on the system and forcing parameters, simulations S1 also show almost regular regimes. In some cases these are much more regular than those obtained in S2 simulations. 

After exploring many different system parameters there appears to be an
explanation for the observed periodicity under certain conditions.  
We observe that in systems where the region of
horizontally excited $H$'s always propagate through most of a unique
cluster, the peaks are more regular.  This is so because for the chain
reaction to take place, a high enough density of
the cluster is needed, and it takes a characteristic time for the cluster to
reach this density after such an event.  If the chain reaction does not
propagate through all the cluster then the time of compression is highly
variable as it depends on the amount of particles that were involved in it,
and also other similar events can take place while the cluster is
compressing in those parts of the cluster that remained dense.  In order for
the explosion to cover most of the $H$ cluster, several conditions must be
fulfilled, particularly that its density is large enough to allow for an uninterrupted
propagation of the horizontal energy.  
Moreover, the concentration of $L$'s inside the cluster needs to be small so that
they do not block the propagating front.  
These conditions can be achieved either by changing the control parameters 
or by having
small geometrical defects such as a small curvature of the box to help the $H$'s
migrate toward the center. Indeed, in S2 simulations, an artificial small curvature
was added and the resulting events are more regular compared to the same simulations done without the small curvature.

{\em Conclusions.} Experiments and numerical simulations of a granular
mixture of grains that differ in their mass density---in a vibrated shallow
box---show horizontal energy peaks characterized by the rapid conversion of
vertical kinetic energy into horizontal one; these events are preceded by
the segregation of the species.  In the segregated state the massive grains
approach a fixed point characterized by a vanishing horizontal energy and
vertical motion in phase with the walls' oscillation.  
The synchronized heavy grains collide between themselves with a tiny momentum
transfer.  As a result, densely packed clusters of the heavy grains develop. 
Eventually, however, a light grain can interpose in the vertical motion of a 
heavy grain and, consequently, the latter acquires
significant horizontal momentum and leaves the fixed point. This can also be triggered by another type of strong fluctuation, like a sudden desynchronization of one $H$ particle due to surface roughness or an energetic incoming $L$ particle. The  subsequent
collisions with neighboring heavy grains transfer a large amount of energy into
the horizontal agitation in the form of a chain reaction, generating an
$E_{Hh}$ peak and an expansion of the heavy particles cluster.
When there is a unique cluster different regimes are reached depending on
the density of the cluster of heavy particles and the relative concentration of light particles
inside it.  When there are too many light particles in the cluster the
horizontal agitation does not involve the entire cluster and the corresponding chain events show no characteristic time or energy scales.  Otherwise, explosions propagate through all
the cluster and show a characteristic intensities and time lapse between successive events.

Simulations and experiments differ in the collision details and,
consequently they do not give the same quantitative results. Although the fixed point exists only for spherical particles, small non-sphericity, as present in the experiments and artificially put in simulations, result in a small horizontal agitation but the energy bursts still take place. The
independence of the phenomenon on the precise details shows that it is
robust, and requires only a large enough mass contrast.  
This phenomenon shows that to correctly describe the collective
dynamics in a confined geometry, the dynamics of individual grains in the
vertical direction is crucial. The revealed link between the small and large scale dynamics should help building hydrodynamic-like models of confined
granular media.

We thank M.G. Clerc for fruitful discussions. This research is supported by
Fondecyt grants 1061112, 1070958, 1090188 and 1100100, and grants 
Anillo ACT 127 and ECOS C07E07.

\end{document}